\documentclass[conference]{IEEEtran}
\IEEEoverridecommandlockouts
\usepackage{cite}
\usepackage{amsmath,amssymb,amsfonts}
\usepackage{url}
\usepackage{algorithmic}
\usepackage{graphicx}
\usepackage{textcomp}
\usepackage{xcolor}
\usepackage{balance}
\usepackage[letterpaper,%
            left=0.75in,right=0.75in,top=0.75in,bottom=1in,%
            footskip=.25in,bindingoffset=0.2in]{geometry}

\graphicspath{{Pictures/}}

\def\BibTeX{{\rm B\kern-.05em{\sc i\kern-.025em b}\kern-.08em
    T\kern-.1667em\lower.7ex\hbox{E}\kern-.125emX}}

\begin{document}

\makeatletter
\newcommand{\linebreakand}{%
 \end{@IEEEauthorhalign}
 \hfill\mbox{}\par
 \mbox{}\hfill\begin{@IEEEauthorhalign}
}
\makeatother

%
\title{Phase Identification of Smart Meters Using a Fourier Series Compression and a Statistical Clustering Algorithm
\thanks{We would like to acknowledge and thank the Post Degree Diploma program, the Work on Campus program, and the Applied Research Centre at Langara College for supporting our research.}
}
\author{
    \IEEEauthorblockN{Jeremy Chiu}
    \IEEEauthorblockA{\textit{Mathematics and Statistics} \\
    \textit{Langara College}\\
    Vancouver, Canada \\
    0000-0002-0737-9055}
   
   \and 
    \IEEEauthorblockN{Albert Wong}
    \IEEEauthorblockA{\textit{Mathematics and Statistics} \\
    \textit{Langara College}\\
    Vancouver, Canada \\
    0000-0002-0669-4352}
    
    \and
    \IEEEauthorblockN{James Park}
    \IEEEauthorblockA{\textit{Mathematics and Statistics} \\
    \textit{Langara College}\\
    Vancouver, Canada \\
    0000-0002-3714-9138}

    \linebreakand

   \IEEEauthorblockN{Joe Mahony}
    \IEEEauthorblockA{\textit{Research and Development} \\
    \textit{Harris SmartWorks}\\
    Ottawa, Canada \\
    JMahony@harriscomputer.com}

\and 
     \IEEEauthorblockN{Michael Ferri}
     \IEEEauthorblockA{\textit{Research and Development} \\
    \textit{Harris SmartWorks}\\
    Ottawa, Canada \\
    mferri@harriscomputer.com}
  
  \and  
    \IEEEauthorblockN{Tim Berson}
    \IEEEauthorblockA{\textit{Research and Development} \\
    \textit{Harris SmartWorks}\\
    Ottawa, Canada \\
    TBerson@harrisutilities.com}
    
}
\maketitle
\begin{abstract}

\textbf{Accurate labeling of phase connectivity in electrical distribution systems is important for maintenance and operations but is often erroneous or missing.  In this paper, we present a process to identify which smart meters must be in the same phase using a hierarchical clustering method on voltage time series data.  Instead of working with the time series data directly, we apply the Fourier transform to represent the data in their frequency domain, remove $98\%$ of the Fourier coefficients, and use the remaining coefficients to cluster the meters are in the same phase.  Result of this process is validated by confirming that cluster (phase) membership of meters does not change over two monthly periods. In addition, we also confirm that meters that belong to the same feeder within the distribution network are correctly classified into the same cluster, that is, assigned to the same phase.}

\end{abstract}

\section{Introduction}
	
Managing an electricity distribution network efficiently requires accurate phase connectivity models\cite{WangYuFoggo}. However, electricity companies usually do not have accurate information of phase connectivity and often require the use of measurement-based phase identification methods.\cite{Hoogsteyn}.

To deliver high-voltage power from the generation station to customers, voltage in the primary distribution circuit is stepped down at a distribution substation. Then through feeders, electricity is distributed to transformers. In North America, power is stepped down again from transformers and distributed to the customers using a three-phase system\cite{WangYuFoggo}. Which phases are used for the customers is often not recorded, and therefore creating a phase identification problem if phase connection information is required for network management tasks.
		
There are many ways in research to tackle this identification problem:
		
\textit{Micro-synchrophasors} - One can use a micro-synchrophasor to measure voltage magnitude and phase angle of a meter \cite{wen2015phase}. The higher the correlation between the voltage magnitude of the substation and that of smart meters, the more accurate the phase labelling. To complete the identification, signal generators are set up at the substations and signal discriminators at the smart meters to accurately identify the phase. This method is quite accurate but expensive as it requires deployment and maintenance of additional equipment and human resources. 
	
\textit{Integer Programming algorithms (\cite{arya,ZhuChowZhang,heidari2021phase,akhijahani2019milp}) } - Phase connection of smart meters are represented as binary variables. Then, integer linear programming methods are used to determine the most-likely phase network.  However, this approach requires a new variable for every new meter, making the problem computationally intensive, especially for feeders with thousands of meters.
		
\textit{Correlation-based method (\cite{6360284,short2012advanced,olivier2018phase})} - Data is first collected over time from the smart meters to be identified. The correlation coefficient is then calculated using voltage time series between two smart meters -- the closer a coefficient is to one, the more likely the pair of smart meters have the same voltage pattern and therefore the same phase. The correlation coefficients are then transformed to a distance measure as input to a clustering algorithm.  The method is logical and seems promising. However, based on results from unpublished research by a project team at Langara College (personal communication), when applied to the data set in this research, this method suffers from issues with a number of performance criteria that we have identified and discussed below.
		
\textit{Constrained k-means clustering } - Voltage time series data is first normalized using standard deviation, then principal component analysis is applied to reduce the data's dimension. A $k$-means clustering algorithm is then used to cluster the smart meters. The phase of each cluster is then identified by solving a minimization problem \cite{WangYuFoggo,olivier2018phase,jayadev2016novel}. 
		
Other phase identification methods proposed include the use of supervised learning models or different types of clustering algorithm, such as spectral clustering \cite{lee2021novel,foggo2018comprehensive,foggo2019improving,ZaragozaRao, wang2017advanced, ma2018phase, blakely2019spectral}. 

In this research, we will take a new approach in the phase identification problem. The central idea is to extract as much information as possible from the voltage time series using a Fourier series compression process. A hierarchical clustering routine is then applied on the compressed data to produce accurate identification.

\section{Research Data Set}	
For this research, we use a voltage data set that was provided by a utility company in the United States, which contains hourly voltage data for a number of smart meters in the month of June and July 2021. The data set also includes the linkage between the smart meters and their associated transformers and feeders. This information is critical for the assessment of appropriateness and accuracy in the clustering results.
	
We removed smart meters with any missing entries from June and July 2021. We then normalize each smart meter by dividing each voltage value by its mean. We chose two of the smaller feeders (Feeder F with 26 smart meters and Feeder D with 55 smart meters) to conduct our research so that we can easily visualize and evaluate the results.

\section{Fourier Compression}
	
Clustering the smart meters using its time series (voltage vs time) is challenging because of its size -- measurements are hourly, so in a month of 30 days, each time series would be in $\mathbb{R}^{720}$.  We reduce the dimension by using a compressed Fourier series, and then cluster the smart meters using the compressed Fourier coefficients. Figure \ref{fig:overview} shows a high level overview of how we use Fourier series to reduce the dimension.

\begin{figure}
     	\centerline{\includegraphics[width=\columnwidth]{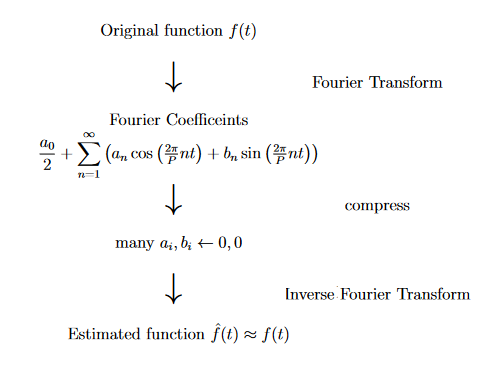}}
		\caption{A high level overview of how we use Fourier series to reduce the dimension of a smart meter.  We performed clustering on the compressed Fourier series.  The functions $f(t)$ and $\hat{f}(t)$ are time series, where $\hat{f}(t)\approx f(t)$.}
		\label{fig:overview}
\end{figure}

	The compression is done as follows.  We represent each smart meter in its frequency domain by applying the Fourier transform to the normalized time series.  Recall the Fourier series (sine-cosine form) representation of a periodic function $f(t)$ is
	\begin{equation}
	     f(t) = \frac{a_0}{2} + \sum_{n=1 }^\infty \left(a_n \cos\left(\tfrac{2\pi}{P} nt \right) + b_n  \sin\left(\tfrac{2\pi}{P} nt \right) \right) ,   
	\end{equation}
	where $a_n,~b_n$ are real coefficients and $P$ is the function's period.  We then delete coefficients that are `small' (either by deleting frequencies that are smaller in magnitude than a predetermined magnitude, or by only keeping a predetermined number of the largest terms), thus giving us a compressed Fourier representation.  We also delete the 0th harmonic $a_0$ because it is constant across all smart meters due to normalization.  In practice, we used 12 Fourier coefficients to represent a month of data, thus reducing the dimension from $\mathbb{R}^{720}$ to $\mathbb{R}^{12}$ (a 98\% reduction in size).

	As demonstrated in Figure \ref{fig:amplitudes}, most of the Fourier coefficients are very small, which suggests the compressed Fourier series could provide a high-accuracy, low-dimension approximation of the time series.  To verify the accuracy of the compression, we obtain an approximate time series by applying the inverse Fourier transform to a compressed Fourier series, and then comparing the approximate time series to the original time series.  Figure \ref{fig:approximate time series} shows approximate time series alongside the original time series -- the general trend of the time series is captured, but the $12$-coefficient approximation does poorly at the spikes.  As Figure \ref{fig:compression error} demonstrates, keeping more coefficients yields better accuracy.  Notice that with about 10\% of the coefficients, we maintain about 90\% accuracy of the time series.  Ultimately, the accuracy of the time series is not too important so long as the clustering results are sensible.
	
	The compression was done in Matlab.  Given a smart meter's time series, we use Matlab's \texttt{fft} function, which returns complex coefficients corresponding to the Fourier series in exponential form.  We convert the complex coefficients into $a_n$ and $b_n$, the real coefficients of the Fourier series in sinusoidal form (we used \texttt{get\_harmonics}\cite{web:getharmonics}).  In practice, a time series in June would be in $\mathbb{R}^{720}$, corresponding to $0\leq t \leq 720$ hours, and so $P=720$.  Matlab's \texttt{fft} would return the complex coefficients $c_{-360}, \ldots, c_{359}$, which we convert to real coefficients, then only keep $a_1, \ldots, a_{360}$ and $b_1, \ldots, b_{360}$ (note $a_{360}$ and $b_{360}$ were computed from $a_{-360}$ and $b_{-360}$).  We then compress by using a mask to set most coefficients to zero.  In practice, we kept $a_n$ and $b_n$ where $n=30, 60, \ldots, 180$ (these coefficients correspond to the large frequencies in Figure \ref{fig:amplitudes}), a total of 12 coefficients.
	
\begin{figure}
		\includegraphics[width=\columnwidth]{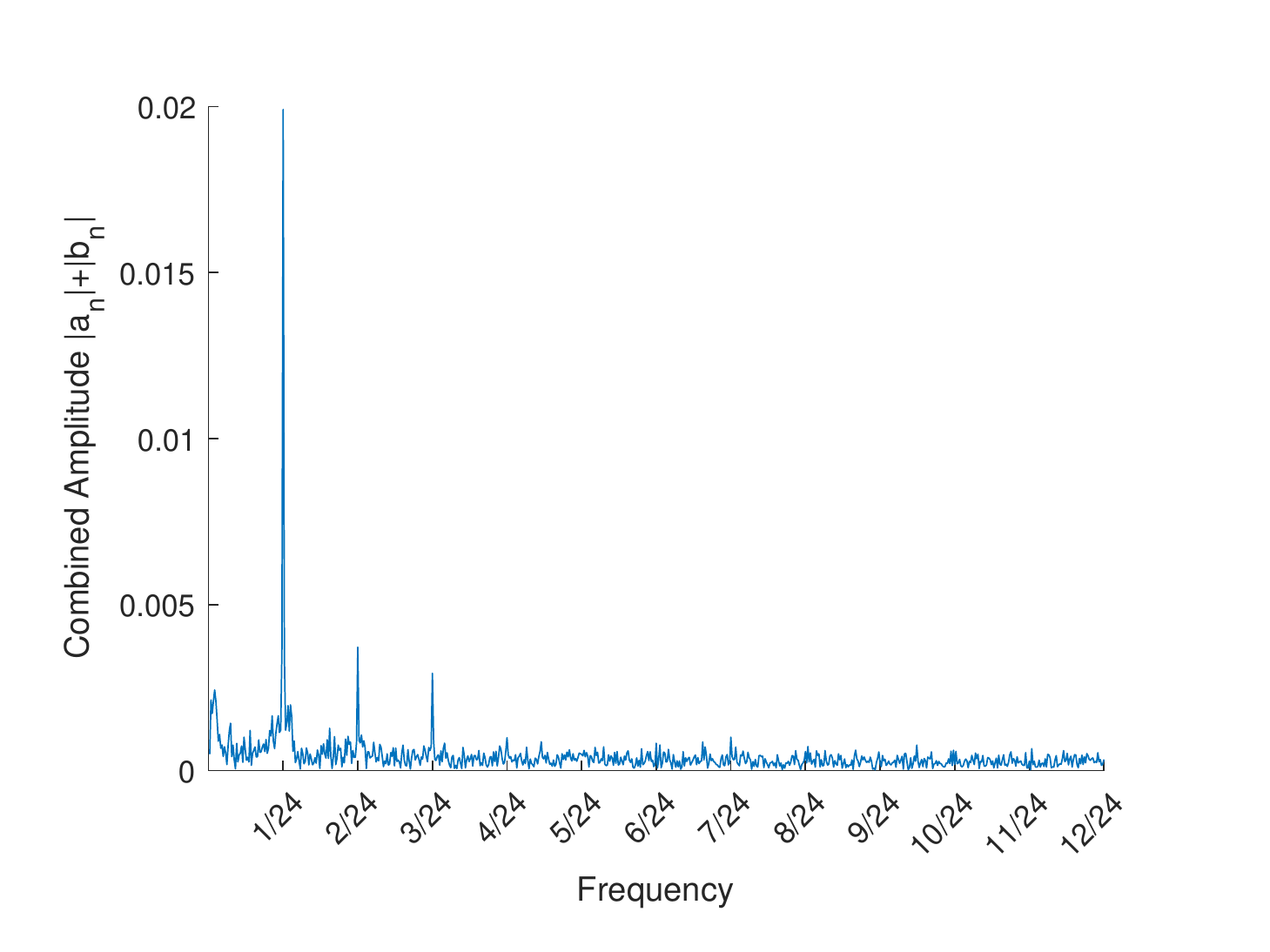}
		\caption{Combined magnitude of the Fourier coefficients $(|a_n|+|b_n|)$ vs frequency.  Notice most of the coefficients are small.  The largest amplitude occur at the frequency $1/24$; this is unsurprising because energy usage follow daily patterns.  
		}
		\label{fig:amplitudes}
\end{figure}

	\begin{figure}
	\includegraphics[width=\columnwidth]{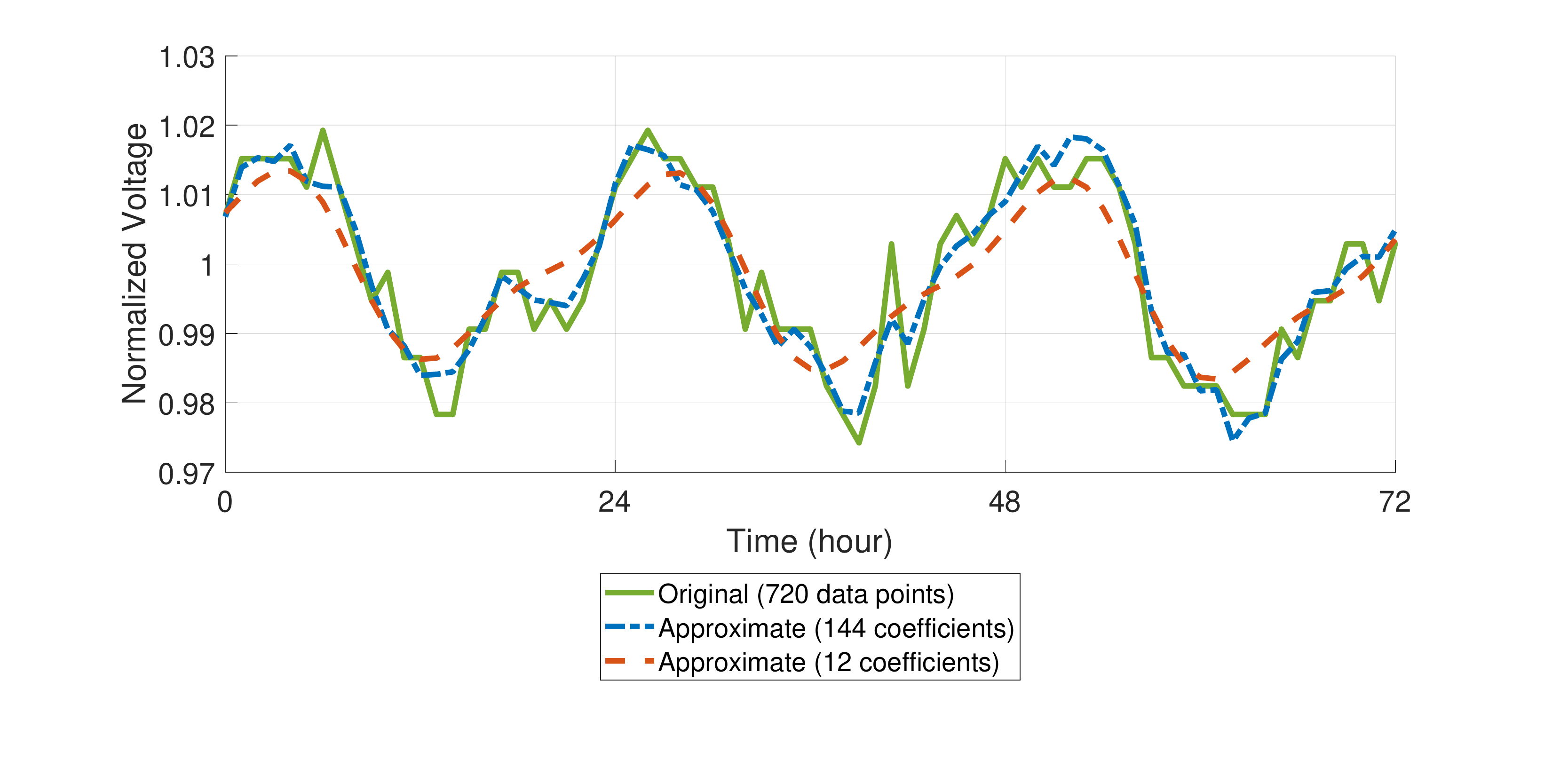}
		
	\caption{Original time series alongside approximate time  series.  The domain was reduced to 3 days for a better viewing rectangle.  The $12$-coefficient approximation does poorly at the spikes, but captures the general trend.  The $144$-coefficient approximation captures most spikes.}
	\label{fig:approximate time series}
\end{figure}
	
	\begin{figure}
		\includegraphics[width=\columnwidth]{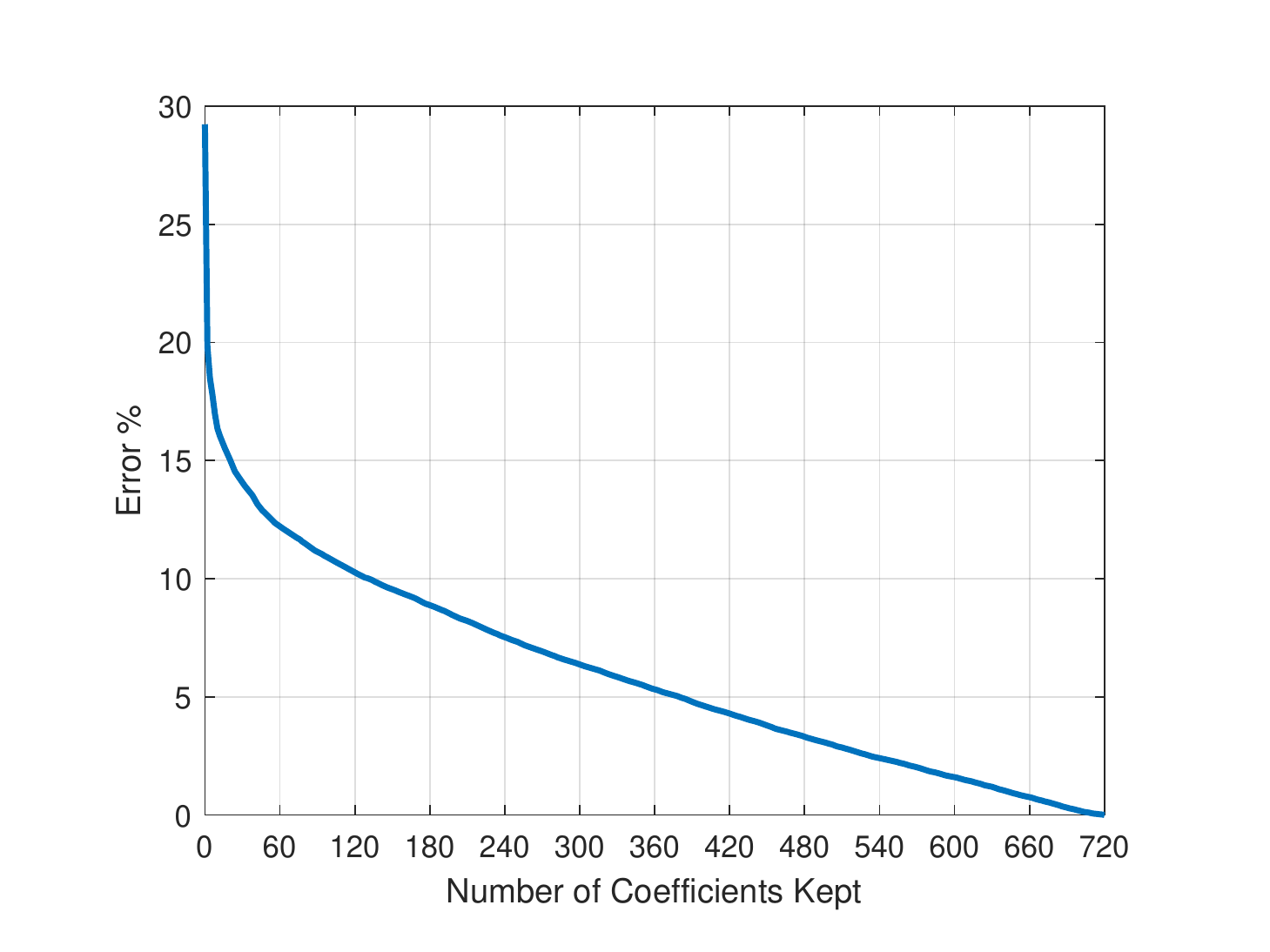}

		\caption{Error percentage is computed as $ \frac{\| y- \hat{y}  \|_2} {\bar{y}}$, where $y$ is the original time series, $\hat{y}$ is the approximate time series, and $\bar{y}$ is the average of $y$ (note $\bar{y}=1$ due to normalization).  The $12$-coefficient approximation has 16\% error.}
		\label{fig:compression error}
	\end{figure}

	
\section{Clustering of Smart Meters}

	

We cluster the set of smart meters in Feeder F (then repeat for Feeder D) using Matlab's Ward hierarchical clustering algorithm \cite{web:ward} with the dimension-reduced Fourier coefficients as input.  Since all smart meters should be in one of the three phases, the number of resulting clusters is set to be three.  Hence, meters clustered together would mean they belong to the same phase.

\section{Validation of Clustering Results}
	
\subsection{Visualizing Clustering Results}

A useful way to visualize the result of clustering a multi-dimensional data set is to somehow ``project" the data set into a two dimensional space. We could then visualize clusters with a scatter diagram in the $xy$-plane. One way to achieve this is to use Matlab's multidimensional scaling technique\cite{web:mds}; given the distance between points, \texttt{mdscale} reconstructs where the points could be in 2D so that the distance is still roughly preserved.  In Figure \ref{fig:MDS D}, we see a visualization of the clustered meters from Feeder D.  Notice that there are clear boundaries between different clusters.

	
Moreover, a hierarchical clustering algorithm such as Ward would allow us to visualize the formation of the clusters hierarchically via a dendogram (Figure \ref{fig:dendogram}). However, it is less useful here because the number of clusters is required to be three.

\begin{figure}
		\centerline{\includegraphics[width=\columnwidth]{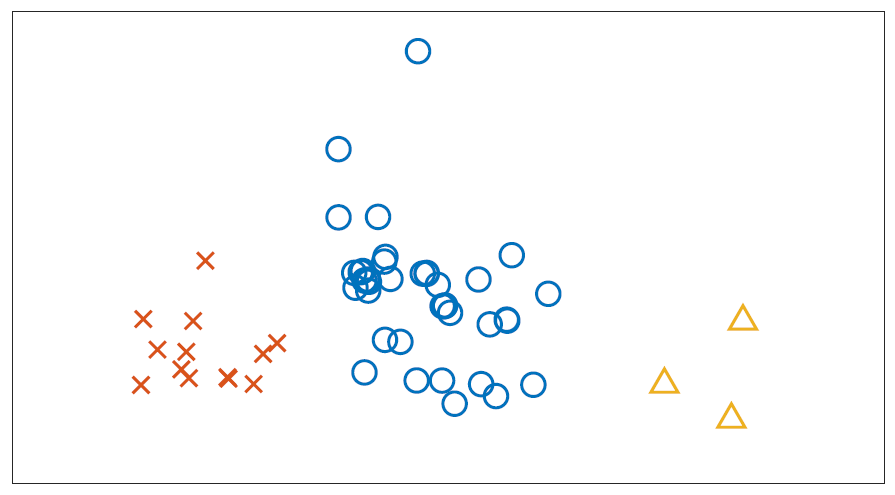}}
		
		\caption{Visualizing the clustering of Feeder D using June 2021 Data via Matlab's \texttt{mdscale} function.}
		\label{fig:MDS D}

\end{figure}

\begin{figure}
	    \centerline{\includegraphics[width=\columnwidth]{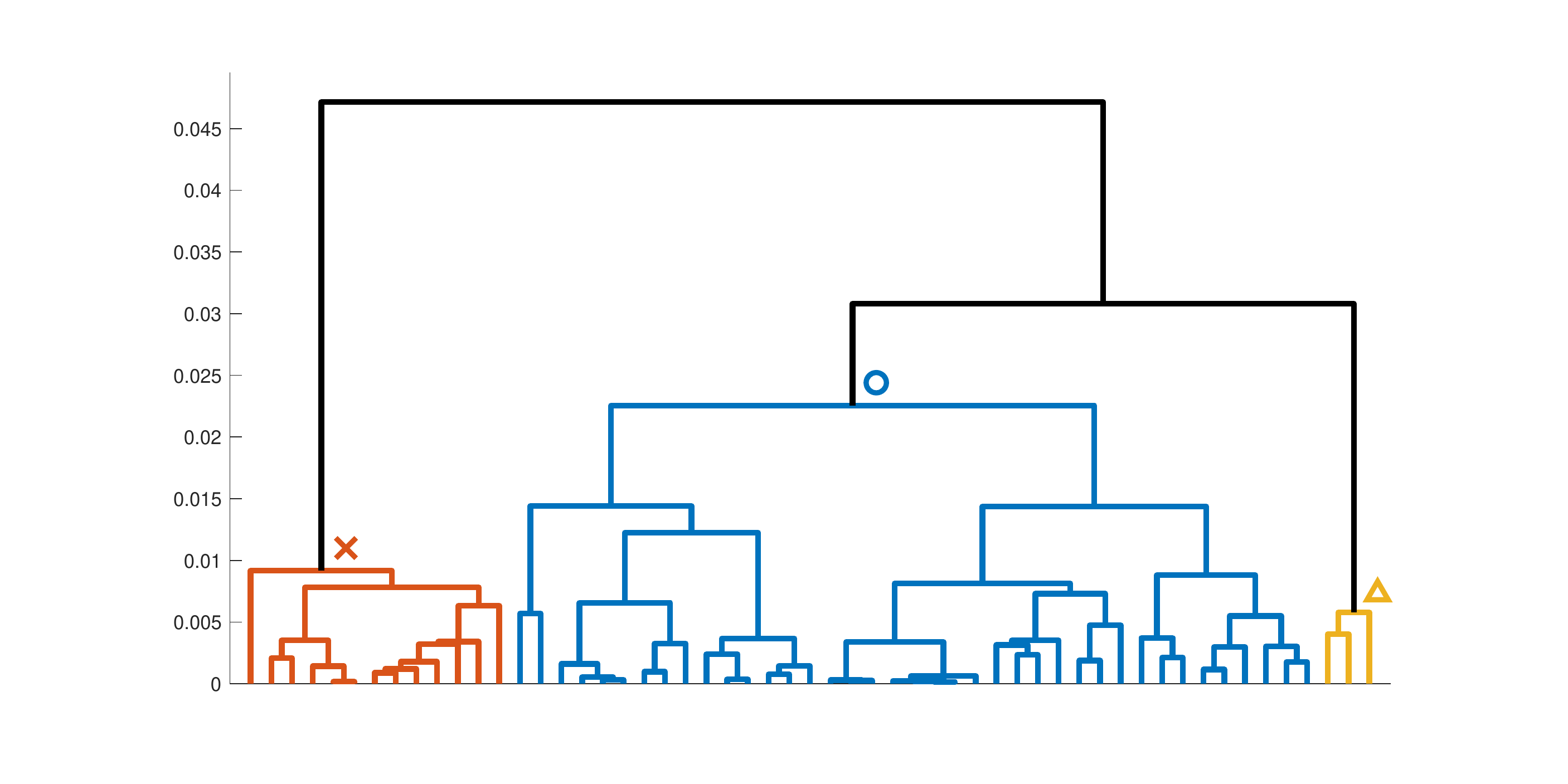}}
	    \caption{Dendogram of clustering Feeder D using data from June 2021.  Dendograms are useful to see how clusters are being formed.}
	    \label{fig:dendogram}
	\end{figure}
	
\subsection{Same Transformer, Same Phase}
	
 Meters within the same transformer must be in the same phase, and thus should be clustered together. We can use this fact to see how well our method performs -- after we cluster the smart meters, each transformer should only have meters of a single phase.  As seen in Tables \ref{Table 1} and \ref{Table 2}, the clustering of Feeder F is almost perfect while that for Feeder D is perfect, giving us hope that this approach has promise.
 

\begin{table}[htb]
    \centering
    
    \caption{Feeder F June 2021 cluster results grouped by transformers.  
    }
         \begin{tabular}{ c || c c c}
         
           &  \multicolumn{3}{c}{Cluster} \\ 
                Transformer & A & B & C \\ \hline
                1 & 1 & 8 & ~ \\ 
                2 & 1 & ~ & ~ \\ 
                3 & 1 & ~ & ~ \\ 
                4 & 1 & ~ & ~ \\ 
                5 & ~ & ~ & 1 \\ 
                6 & 4 & ~ & ~ \\ 
                7 & ~ & ~ & 1 \\ 
                8 & ~ & ~ & 1 \\ 
                9 & ~ & ~ & 1 \\ 
                10 & ~ & ~ & 1 \\ 
                11 & 5 & ~ & ~ \\  \hline
                Total & 13 & 8 & 5 \\ 
            \end{tabular} 
    \bigskip

    \label{Table 1}
\end{table}

\begin{table}[htb]
    \centering
    
        \caption{Feeder D June 2021 cluster results grouped by transformers.  }
    \label{Table 2}
          \begin{tabular}{ c || c c c}
         
           &  \multicolumn{3}{c}{Cluster} \\ 
       Transformer & A & B & C \\ \hline
            1 & 1 & ~ & ~ \\ 
            2 & 1 & ~ & ~ \\ 
            3 & 1 & ~ & ~ \\ 
            4 & 1 & ~ & ~ \\ 
            5 & 2 & ~ & ~ \\ 
            6 & 1 & ~ & ~ \\ 
            7 & ~ & ~ & 1 \\ 
            8 & 1 & ~ & ~ \\ 
            9 & 1 & ~ & ~ \\ 
            10 & 4 & ~ & ~ \\ 
            11 & 1 & ~ & ~ \\ 
            12 & ~ & ~ & 1 \\ 
            13 & ~ & ~ & 1 \\ 
            14 & 1 & ~ & ~ \\ 
            15 & ~ & 1 & ~ \\ 
            16 & 1 & ~ & ~ \\ 
            17 & 2 & ~ & ~ \\ 
            18 & 1 & ~ & ~ \\ 
            19 & 1 & ~ & ~ \\ 
            20 & ~ & 1 & ~ \\ 
            21 & ~ & 1 & ~ \\ 
            22 & ~ & 3 & ~ \\ 
            23 & ~ & 1 & ~ \\ 
            24 & ~ & 2 & ~ \\ 
            25 & 2 & ~ & ~ \\ 
            26 & ~ & 1 & ~ \\ 
            27 & 1 & ~ & ~ \\ 
            28 & 1 & ~ & ~ \\ 
            29 & ~ & 1 & ~ \\ 
            30 & 1 & ~ & ~ \\ 
            31 & 2 & ~ & ~ \\ 
            32 & 1 & ~ & ~ \\ 
            33 & 4 & ~ & ~ \\ 
            34 & 1 & ~ & ~ \\ 
            35 & 1 & ~ & ~ \\ 
            36 & ~ & 1 & ~ \\ 
            37 & 2 & ~ & ~ \\ 
            38 & 3 & ~ & ~ \\ 
            39 & ~ & 1 & ~ \\ \hline
            Total & 39 & 13 & 3 \\ 
            \end{tabular} 
        \bigskip

\end{table}

	

\subsection{Stability Over Time}
	
Physically, meters do not change phase over time. Therefore, for the clustering (assignment of phase) to be meaningful, the result should not change over time. 

To evaluate results from this research, we performed cluster analysis on two different time periods (June 2021 and July 2021) on Feeder F and D, then checked for inconsistent results.  Any meter that changed phases (clusters) are considered time unstable. Note that the labels from the clustering (A, B, and C) are arbitrary, and so we use a cross tabulation of the two clustering results to see how meters are assigned in the clustering processes.  Table \ref{Table 3} shows that the clustering from June to July is stable. All 13 meters assigned to Cluster A in June are also assigned in the same cluster in July; the same is true for Clusters B and C.

\begin{table}[!ht]
    \centering
    
    \caption{Feeder F cluster results in June and July 2021.}
    
    \begin{tabular}{c c|c c c|c}
        & ~ & ~ & July & ~ & ~ \\ 
         & ~ & A & B & C & Total \\ \hline
        & A & 13 & ~ & ~ & 13 \\ 
        June & B & ~ & 8 & ~ & 8 \\ 
        & C & ~ & ~ & 5 & 5 \\ \hline 
        & Total  & 13 & 8 & 5 & 26 \\ 
    \end{tabular}
    
    \label{Table 3}
\end{table}

The same can be said about the stability of clustering Feeder D using our approach (Table \ref{Table 4}).

\begin{table}[!ht]
    \centering
    
        \caption{Feeder D cluster results in June and July 2021.}

    \begin{tabular}{c c|c c c|c}
        & ~ & ~ & July & ~ & ~ \\ 
         & ~ & A & B & C & Total \\ \hline
        & A & 39 & ~ & ~ & 39 \\ 
        June & B & ~ & 13 & ~ & 13 \\ 
        & C & ~ & ~ & 3 & 3 \\ \hline 
        & Total  & 39 & 13 & 3 & 55 \\ 
    \end{tabular}
    
    \label{Table 4}
\end{table}

\section{Future Work}

While the above results look very promising, we have not applied this approach to a larger feeder (say with over 300 meters), or to a data set with multiple feeders. We suspect, due to the increased likelihood of data related issues, that the results may not be as ``perfect" as we have seen so far.

To advance our research, the approach would be applied to a larger data set with multiple feeders. The same approach should also be applied to a data set with several months; clustering could be done month by month, or with several months combined.  Considerations should also be given to use this approach to cluster a subset of the data set and, after  the validation process as outlined above, using the cluster labels for the development of a supervised learning model for the classification of other meters. 
	
\section{Conclusion}	
	
In this research, we have applied a novel method of approximating a time series with its Fourier series.  We then used hierarchical clustering methods on the dimension-reduced data. The major application of this approach is in the phase identification of smart meters in a network environment.

Results from two small data sets using this approach show significant promise as they passed two important tests: same assignment for meters in the same transformer and stability of assignment over time. The application of this approach to a larger data set with multiple feeders would therefore be a worthwhile exercise.  
	
	
	\bibliographystyle{siam}
	\bibliography{Phase_Identification_of_Smart_Meters_using_a_Fourier_Series_Approximation_Paper}
	
\end{document}